\documentclass[11pt]{article} 
\pdfoutput=1

\usepackage[letterpaper,margin=1in]{geometry}
\usepackage{pdflscape}

\usepackage[svgnames]{xcolor} 
\definecolor{gublue}{RGB}{4, 30, 66}

\usepackage{comment}
\usepackage[onehalfspacing]{setspace}
\usepackage[ 
	colorlinks = true , 
	urlcolor   = gublue , 
	linkcolor  = black ,
	citecolor  = gublue 
	]{hyperref}
\usepackage{fontawesome}

\usepackage{titlesec}
\titleformat*{\section}{\normalfont\sffamily\Large\sc\color{gublue}} 
\titleformat*{\subsection}{\normalfont\sffamily\large\sc\color{gublue}} 

\usepackage{titletoc}
\dottedcontents{section}[0em]{\sc}{2.9em}{1pc}

\usepackage{amsmath}
\usepackage{amsfonts}
\usepackage{amssymb}
\usepackage{amsthm}
 
\usepackage[authoryear,sort&compress,round]{natbib}
\usepackage[nottoc,numbib]{tocbibind} 

\usepackage{enumerate}

\usepackage{afterpage} 
\usepackage[capposition=top,footfont=small,subfloatrowsep=none]{floatrow}

\usepackage{graphicx}
\usepackage[position=bottom]{subfig} 

\usepackage{makecell}
\usepackage{booktabs}
\usepackage{multirow}
\usepackage{rotating}
\usepackage{array}
\newcolumntype{L}[1]{>{\raggedright\let\newline\\\arraybackslash\hspace{0pt}}m{#1}}
\newcolumntype{C}[1]{>{\centering}m{#1}}
\usepackage{caption} 
\usepackage{siunitx}
\usepackage{appendix}

\usepackage{chngcntr}

\title{Teacher turnover in Rwanda}
\date{September 28, 2020}
\author{Andrew Zeitlin\thanks{McCourt School of Public Policy, Georgetown University. Email: \texttt{andrew.zeitlin@georgetown.edu.}}\,\,%
\thanks{For guidance and input into this project, I am grateful to Dr. Ir\'en\'ee Ndayambaje, James Ngoga, and Johnson Ntagaramba at the Rwanda Education Board, to Dr. Isaac Munyakazi and Samuel Mulindwa at the Ministry of Education, to Leodomir Mfura, Sophie Mushimiyimana, and Doug Kirke-Smith at Innovations for Poverty Action, and to Jonathan Bower at the International Growth Centre.  Grace Han, Showroop Pokhrel, Aruj Shukla, and Ana Gabriela Tamayo Alatriste provided excellent research assistance. All errors are my own.}}

\begin{document}

{
\normalfont\sffamily\Large\sc\color{gublue}
\maketitle
}
\thispagestyle{empty}

\begin{abstract}
Despite widely documented shortfalls of teacher skills and effort, there is little systematic evidence of rates of teacher turnover in low-income countries. I investigate the incidence and consequences of teacher turnover in Rwandan public primary schools over the period from 2016--2019. To do so, I combine the universe of teacher placement records with student enrollment figures and school-average Primary Leaving Exam scores in a nationally representative sample of 259 schools. Results highlight five features of teacher turnover. First, rates of teacher turnover are high: annually, 20 percent of teachers separate from their jobs, of which 11 percent exit from the public-sector teaching workforce. Second, the burden of teacher churn is higher in schools with low learning levels and, perhaps surprisingly, in low pupil-teacher-ratio schools. Third, teacher turnover is concentrated among early-career teachers, male teachers, and those assigned to teach Math. Fourth, replacing teachers quickly after they exit is a challenge; 23 percent of exiting teachers are not replaced the following year. And fifth, teacher turnover is associated with subsequent declines in learning outcomes.  On average, the loss of a teacher is associated with a reduction in learning levels of 0.05 standard deviations. In addition to class-size increases, a possible mechanism for these learning outcomes is the prevalence of teachers teaching outside of their areas of subject expertise: in any given year, at least 21 percent of teachers teach in subjects in which they have not been trained. Taken together, these results suggest that the problem of teacher turnover is substantial in magnitude and consequential for learning outcomes in schools. 
\end{abstract}

\setcounter{page}{0}
\clearpage

\clearpage 
\section{Introduction}\label{s:intro}

Government capacity to recruit, motivate, and retain an effective teaching workforce is crucial to the goal of raising student learning outcomes throughout the world.  On average, teacher salaries comprise as much as three quarters of average education budgets in low-income countries \citep{BruMinRak03wb}, and it is estimated that developing countries will need to recruit an additional 68.8 million teachers by 2030 to meet the Sustainable Development Goal of universal primary and secondary education \citep{unesco16goals}.  Meanwhile, there is substantial variation in the cognitive \citep{CheFriRoc14aer1,Buh16wp} and non-cognitive \citep{Jac18jpe} skills that teachers help students to achieve.  These facts suggest that the challenge of developing effective personnel systems is central to the mission of educational quality \citep{FinOlkPan17hbk}.

Two dimensions of this personnel problem have been widely recognized:  teacher skills and teacher effort. The World Bank's Service Delivery Indicators surveys estimate that, across a sample drawn from seven sub-Saharan African Countries, only 7 percent of primary school teachers possess sufficient curricular knowledge to teach languages at the fourth-grade level, and only 68 percent of primary school teachers possess sufficient knowledge for teaching in mathematics \citep{Bol17jep}.  These surveys also reveal an effort gap:  unannounced visits revealed teachers to be absent from the school 23 percent of the time, and absent from the classroom 44 percent of the time, corroborating earlier findings from across the developing world \citep{ChaHamKreMurRog06jep}.  While the severity and specific manifestation of knowledge and effort constraints will vary across countries and systems, there is little doubt that these are important considerations for the design of any personnel system.

In this paper, I highlight a third dimension of the challenge of providing effective human resources for developing-country systems:  \emph{teacher turnover}. The separation of teachers from their posts can occur because they transfer or because they drop out of the teaching workforce altogether.  Teacher turnover affects the production of student learning not only indirectly, through the stock of teachers' skill and motivation, but also directly, because it introduces frictions in the allocation of teachers to schools. This misallocation problem is not well captured in the traditional emphasis on aggregate pupil-teacher ratios.  At the school level, head teachers faced with staffing shortages may be forced to allocate teachers to subjects outside of their areas of specialty, resulting in a patchwork of teaching assignments that adversely affects learning outcomes.  

Teacher turnover may have benefits alongside these costs.  For example, when low-performing teachers are disproportionately likely to exit a school system, the overall quality of teaching can improve \citep{DeeWyc15jpam,Rot15aer}.  But designing human resource systems to encourage such beneficial exit patterns is difficult, to the extent that higher value-added teachers have better outside options.\footnote{This appears to be the case in the U.S.: see \citet{FigKen91jpubec}.}  Moreover, the consequences of teacher exits will depend on when, in the life cycle of their careers, teachers are most likely to leave.  Retirements may have very different consequences than early-career exits.  Finally, the impacts of teacher turnover will depend on whether, how quickly, and with what quality of teacher the school system is able to replace those who leave.  Because the supply of teachers and the incidence of exits may differ across locations, the net effect of this phenomenon may vary across space.

In spite of the potential importance of teacher turnover to the performance of education systems, there exists little systematic evidence pertaining to this phenomenon in low-income countries in general, and in sub-Saharan Africa in particular. What evidence we do have suggests that in high- and middle-income settings, teacher exits are comparatively low.  In the United States, approximately eight percent of teachers are estimated to have exited annually in 2012--13, with an additional eight percent switching schools \citep{GolTaiRid14nces}, but annual exit rates among early-career teachers appear to be substantially lower \citep{RauGra15nces}.  In the U.K., a mere 2.7 percent of teachers exit the system annually, with a further 7.8 percent transferring schools.  In Chile, \citet{BehTinTodWol16jole} report that fewer than 3 percent of teachers exit the municipal school system annually.  Nationally representative studies of this phenomenon are rare in low-income countries, and in Sub-Saharan Africa in particular.\footnote{For example, the recent World Development Report on education mentions teacher retention only once, and provides no descriptive information on teacher retention rates \citep{wdr2018}.} This dearth of evidence may persist in part because many education management information systems are not set up to track individual teachers over time, particularly in contexts where payrolls are managed by Ministries of Finance rather than Ministries of Education.

This paper addresses that gap by combining several distinct administrative datasets from Rwanda.  Using teacher deployment data from the years 2016--2019, I construct a panel of teachers over time.  Student enrollment data from the year 2019, assuming the relative stability of enrollment figures over time, allow me to shed light on how the incidence of teacher exits varies with pupil-teacher ratios.  I combine the above sources with data on learning outcomes, measured by the nationally administered Primary 6 exam, in a representative sample of schools, for the years 2016--2018.  The ability to match teacher turnover to school learning outcomes allows me to shed light on how the incidence of teacher exits varies by school learning levels, and to provide observational evidence on the learning consequences of those exits, 

I draw five lessons from these data. 
First, rates of teacher turnover are high.  On average, 11 percent of primary teachers exit the public-sector teaching workforce in a given year. A further 8 percent transfer to another school within the same district, and 1 percent of teachers transfer across districts annually.  Consequently 20 percent of teachers separate from their jobs on average in any given year.
Second, the burden of teacher churn is higher in schools with low learning levels.  Teacher churn is also higher in low pupil-teacher-ratio schools, a finding consistent with district redistributive policies being an important determinant of teacher churn.
Third, not all teachers are equally likely to leave.  Teacher turnover is concentrated among early-career teachers.  Female teachers are less likely to exit, whereas those teachers assigned to teach Math disproportionately exit the education system altogether.
Fourth, replacing teachers quickly after they exit is a challenge; when teachers exit a school, districts are often delayed in replacing them. On average, following the loss of a teacher, districts were able to provide schools with replacements by the next year only 77 percent of the time; put differently, schools that have lost a teacher remain with fewer total teachers the next year in 23 percent of cases.  In some districts replacement rates below 60 percent were observed.
And fifth, teacher turnover is associated with subsequent declines in learning outcomes, though caution is required in interpreting these estimates as causal.  On average, the loss of a teacher is associated with a reduction in learning levels of 0.05 standard deviations, with the strongest associations between teacher exits and subsequent learning losses concentrated among early-career teachers and teachers assigned to teach English and Kinyarwanda.  In addition to class-size increases, a possible mechanism for these learning outcomes is the prevalence of teachers teaching outside of their areas of subject expertise: in any given year, I estimate that at least a lower bound of 21.7 percent of teachers teach in subjects in which they have not been trained.

The remainder of this paper is structured as follows.  In Section \ref{s:context}, I describe the institutional context governing teacher management policy in Rwanda.  Section \ref{s:data} describes the data employed.  Results are presented in Section \ref{s:results}.  Section \ref{s:conclusions} concludes with implications for policy and future research.

\section{Institutional context}\label{s:context}

The governance structure for the hiring, deployment, management, and dismissal of teachers in Rwanda involves both national and district-level organizations.  Under Presidential Order Number 24/01 of 2016, the recruitment of teachers in the period studied here was undertaken in parallel by the 30 districts of Rwanda, with most of the responsibility for management of teachers occurring at the district level as well.  Oversight was provided by the Rwanda Education Board (REB), the relevant implementing agency of the Ministry of Education (MINEDUC). Districts' ability to recruit was subject to their available hiring lines, which were approved centrally in a process led by REB and MINEDUC. In general, teachers were required to possess specific qualifications in order to be able to apply for a post; in the case of primary education, the required degree was a secondary-equivalent Teacher Training College Certificate.\label{text:ttcfix}  However, the governing statute did provide for hires that do not have these qualifications, subject to the requirement that they complete a transitional training provided by the Ministry of Education within three years of their appointment.\footnote{While systematic figures on contract teachers in Rwanda are rare, \citet{BenNta08report} report that approximately 5 percent of teachers in 2008 were contract teachers hired by District Education Offices.  In the Government's Education Sector Strategic Plan for the period 2013/14--2017/18, the Ministry of Education adopted the aim of raising the proportion of formally qualified teachers \citep{mineduc2013essp}.}  Teacher transfers within the districts of their posting were relatively easily accomplished, and may be initiated either by the teacher themselves or by the district education office.  The process for transfers across districts was more arduous, requiring a letter to be sent from the originating district's mayor to the mayor of the desired destination district.  Dismissal procedures were governed by general civil service statutes.   

At the primary level in particular, teacher salaries were comparatively low during the period studied.\footnote{The Government of Rwanda has since embarked on a path of raising teacher salaries aimed to improve competitiveness of jobs in this sector.  However, these policies were not in place during the period studied in this paper.} Base pay for a primary teacher stood at RWF 32,500 per month in 2012, with a bonus of RWF 12,500 paid from capitation grant funds, for a total of approximately USD 70 per month at 2012 nominal exchange rates \citep{ipar2012observatory}.   In PPP-adjusted terms,  primary teacher salaries rank below those of Tanzania and Uganda, among other neighbors \citep{wb_gor_19_drivers}.  Compared to teachers with secondary equivalent educations in those countries as reported in \citet{EvYuaFil20cgd}, Rwandan primary teacher salaries are slightly lower than those in Uganda less than a quarter of those in Tanzania. 
Domestically, this lack of competitiveness is true in relative as well as absolute terms.  Using data from the 2017 Rwanda Labour Force Survey, \citet{LeaOziSerZei19stars} find that only 37 percent of TTC graduates were in teaching jobs, with a further 15 percent of TTC graduates in other forms of salaried employment.  Salaried TTC graduates outside the education sector earned an average premium of 30 percent over their peers in teaching jobs.  Low salaries for primary school teachers are potentially an important cause of turnover rates observed in this sector. 

Under the Education Sector Strategic Plan that governed policy under the period studied, a variety of complementary actions were taken to support teacher recruitment, morale, and retention \citep{mineduc2013essp}.  These included the use of a Teacher Development Fund to provide laptops to approximately 1,000 teachers per year, and other non-monetary incentives such as cows and motorbikes. Teachers could also access loans through Mwalimu SACCO.  However, the numbers of teachers reached by such policies are limited, and they were not targeted systematically on the basis of the learning outcomes delivered by teachers.\footnote{In some cases, head teachers were asked to nominate effective teachers for recognition, with the district then choosing a subset of these teachers for awards. In interviews, some teachers reported rotating these nominations among their staff as a response to pressure to treat staff equitably.}  These factors may limit the extent to which such policies focus resources on the retention of those teachers who contribute the most to student learning outcomes, a key factor in whether teacher turnover serves to improve overall learning outcomes \citep{Rot15aer}.

\section{Data}\label{s:data}

This paper combines three data sources---each an administrative dataset received from REB for this purpose.  Most central are administrative data maintained by the Department of Teacher Development and Management, containing teacher placements from 2016--2019.  Over this period, these data contain teachers' National IDs, allowing them to be linked reliaby across rounds of the data.  They also contain a range of teacher characteristics, including ages, lengths of tenure, genders, and subjects taught. Second, a parallel dataset provides pupil enrollment figures for primary schools.  And third, for a representative sample of 295 schools drawn in each of Rwanda's 30 districts, I employ data on average student P6 results for the years 2016--2018.  We match these schools to school names in the TDM teacher listing in order to be able to correlate patterns of employment with exam outcomes.  This provides us with a \emph{matched sample} of schools for which I have teacher assignments, student enrollments and exam data.

\begin{table}[!hbtp]
\caption{Teacher characteristics by year of employment}
\label{t:teacherchars}

\begin{footnotesize}
\begin{center}
\begin{tabular}{l *{4}{S}}
\toprule
 & \multicolumn{4}{c}{Year} \\ 
\cmidrule(lr){2-5}
\multicolumn{1}{c}{\text{ }} & \multicolumn{1}{c}{\text{2016}} & \multicolumn{1}{c}{\text{2017}} & \multicolumn{1}{c}{\text{2018}} & \multicolumn{1}{c}{\text{2019}}\\
\midrule
\multicolumn{5}{l}{\emph{Panel A.  All teachers}}  \\ 
\addlinespace[1ex] \multirow[t]{ 2}{0.2\textwidth}{Tenure} &     10.67 &     11.54 &     12.15 &     12.09 \\ 
 & (9.62)  & (9.61)  & (9.55)  & (9.68)  \\ 
\addlinespace[1ex] \multirow[t]{ 2}{0.2\textwidth}{Age} &     36.56 &     36.87 &     37.31 &     37.09 \\ 
 & (8.86)  & (8.93)  & (9.01)  & (9.20)  \\ 
\addlinespace[1ex] \multirow[t]{ 2}{0.2\textwidth}{Female} &      0.48 &      0.46 &      0.46 &      0.48 \\ 
 & (0.50)  & (0.50)  & (0.50)  & (0.50)  \\ 
\addlinespace[1ex] \multirow[t]{ 2}{0.2\textwidth}{Teaching English} &      0.29 &      0.28 &      0.29 &      0.26 \\ 
 & (0.46)  & (0.45)  & (0.45)  & (0.44)  \\ 
\addlinespace[1ex] \multirow[t]{ 2}{0.2\textwidth}{Teaching Kinyarwanda} &      0.30 &      0.30 &      0.29 &      0.26 \\ 
 & (0.46)  & (0.46)  & (0.45)  & (0.44)  \\ 
\addlinespace[1ex] \multirow[t]{ 2}{0.2\textwidth}{Teaching Math} &      0.32 &      0.32 &      0.31 &      0.28 \\ 
 & (0.47)  & (0.46)  & (0.46)  & (0.45)  \\ 
\addlinespace[1ex] \multirow[t]{ 2}{0.2\textwidth}{Teaching Science} &      0.12 &      0.13 &      0.13 &      0.12 \\ 
 & (0.32)  & (0.34)  & (0.34)  & (0.32)  \\ 
\addlinespace[1ex] \multirow[t]{ 2}{0.2\textwidth}{Teaching Social Studies} &      0.20 &      0.18 &      0.18 &      0.19 \\ 
 & (0.40)  & (0.39)  & (0.39)  & (0.40)  \\ 
\addlinespace[1ex] \multicolumn{5}{l}{\emph{Panel B.  Teachers in sample schools}}  \\ 
\addlinespace[1ex] \multirow[t]{ 2}{0.2\textwidth}{Tenure} &     10.76 &     11.48 &     12.27 &     12.25 \\ 
 & (9.64)  & (9.58)  & (9.55)  & (9.63)  \\ 
\addlinespace[1ex] \multirow[t]{ 2}{0.2\textwidth}{Age} &     36.61 &     36.85 &     37.51 &     37.33 \\ 
 & (8.85)  & (8.93)  & (9.05)  & (9.15)  \\ 
\addlinespace[1ex] \multirow[t]{ 2}{0.2\textwidth}{Female} &      0.50 &      0.45 &      0.47 &      0.49 \\ 
 & (0.50)  & (0.50)  & (0.50)  & (0.50)  \\ 
\addlinespace[1ex] \multirow[t]{ 2}{0.2\textwidth}{Teaching English} &      0.30 &      0.29 &      0.28 &      0.27 \\ 
 & (0.46)  & (0.45)  & (0.45)  & (0.44)  \\ 
\addlinespace[1ex] \multirow[t]{ 2}{0.2\textwidth}{Teaching Kinyarwanda} &      0.30 &      0.30 &      0.28 &      0.24 \\ 
 & (0.46)  & (0.46)  & (0.45)  & (0.43)  \\ 
\addlinespace[1ex] \multirow[t]{ 2}{0.2\textwidth}{Teaching Math} &      0.32 &      0.32 &      0.30 &      0.26 \\ 
 & (0.47)  & (0.47)  & (0.46)  & (0.44)  \\ 
\addlinespace[1ex] \multirow[t]{ 2}{0.2\textwidth}{Teaching Science} &      0.12 &      0.14 &      0.13 &      0.11 \\ 
 & (0.32)  & (0.35)  & (0.33)  & (0.32)  \\ 
\addlinespace[1ex] \multirow[t]{ 2}{0.2\textwidth}{Teaching Social Studies} &      0.20 &      0.20 &      0.18 &      0.21 \\ 
 & (0.40)  & (0.40)  & (0.38)  & (0.41)  \\ 
\addlinespace[1ex] All teachers  & \multicolumn{1}{c}{   37,932}  & \multicolumn{1}{c}{   37,833}  & \multicolumn{1}{c}{   36,277}  & \multicolumn{1}{c}{   40,520}  \\ 
Sample teachers  & \multicolumn{1}{c}{    4,015}  & \multicolumn{1}{c}{    4,085}  & \multicolumn{1}{c}{    3,852}  & \multicolumn{1}{c}{    4,246}  \\ 
Sample schools  & \multicolumn{1}{c}{246}  & \multicolumn{1}{c}{250}  & \multicolumn{1}{c}{239}  & \multicolumn{1}{c}{248}  \\ 
\addlinespace[1ex] 
\bottomrule
\end{tabular}

\end{center}
\vskip-2ex
\floatfoot{
    \textsc{Notes}---Standard deviations in parentheses.   Teachers in sample schools refer to those who, in the relevant year, were found working in the matched sample of schools for which REB exam data were provided, and which could be matched to the teacher placement dataset.
}

\end{footnotesize}
\end{table}

I summarize information on teachers in Table \ref{t:teacherchars}, where Panel A presents estimates based on the census of teachers in REB data.  Total numbers of teachers have fluctuated across the four years under study, but teacher attributes are generally stable.  Teachers have an average of around 10 years of experience, with substantial variation.  Their mean age is approximately 36.  47 percent of teachers are female.  And coverage of subjects is, for the most part, evenly spread across teachers, with one exception:  the relatively low number who teach science suggests that schools disproportionately give one (or a small number of) teacher(s) the responsibility of teaching science.  In Panel B, I replicate these figures using only teachers who are employed in that year at one of the sample of schools for which I have exams data. The share of exam-data schools that can be matched to teacher placement data varies by year, with a maximum of 250. Schools are matched across years by name and district, as there is no unique school identifier in these data.  Consequently, in a small number of cases a school is not observed in consecutive years; in those cases, teacher transfer outcomes will be treated as missing.\label{text:missingdata}  The similarity of all figures in Panel B to those in Panel A provides evidence of the representative nature of the matched sample relative to the national distribution of teachers.

\begin{table}[!hbtp]
\caption{Exam outcomes and teacher placement in sample schools}
\label{t:schoolchars}

\begin{footnotesize}
\begin{center}
\begin{tabular}{l *{4}{S}}
\toprule
 & \multicolumn{4}{c}{Year} \\ 
\cmidrule(lr){2-5}
\multicolumn{1}{c}{\text{ }} & \multicolumn{1}{c}{\text{2016}} & \multicolumn{1}{c}{\text{2017}} & \multicolumn{1}{c}{\text{2018}} & \multicolumn{1}{c}{\text{2019}}\\
\midrule
\multicolumn{5}{l}{\emph{Panel A. Examinations data}}  \\ 
\addlinespace[1ex] \multirow[t]{ 2}{0.2\textwidth}{Average exam score} &     34.16 &     34.94 &     36.82 &         . \\ 
 & (3.97)  & (4.09)  & (3.30)  &  \\ 
\addlinespace[1ex] \multicolumn{5}{l}{\emph{Panel B. Teacher placement data}}  \\ 
\addlinespace[1ex] \multirow[t]{ 2}{0.2\textwidth}{Number of teachers} &     20.69 &     20.52 &     21.01 &     21.98 \\ 
 & (    11.32)  & (    10.84)  & (    10.94)  & (    11.29)  \\ 
\addlinespace[1ex] \multirow[t]{ 2}{0.2\textwidth}{Teaching English} &      6.00 &      5.44 &      5.95 &      5.71 \\ 
 & (3.81)  & (3.66)  & (3.94)  & (3.59)  \\ 
\addlinespace[1ex] \multirow[t]{ 2}{0.2\textwidth}{Teaching Kinyarwanda} &      6.04 &      5.83 &      6.03 &      5.30 \\ 
 & (3.82)  & (3.92)  & (4.09)  & (3.72)  \\ 
\addlinespace[1ex] \multirow[t]{ 2}{0.2\textwidth}{Teaching Math} &      6.35 &      6.21 &      6.39 &      5.69 \\ 
 & (4.00)  & (3.96)  & (4.09)  & (3.82)  \\ 
\addlinespace[1ex] \multirow[t]{ 2}{0.2\textwidth}{Teaching Science} &      2.30 &      2.66 &      2.71 &      2.36 \\ 
 & (1.94)  & (2.17)  & (2.13)  & (1.99)  \\ 
\addlinespace[1ex] \multirow[t]{ 2}{0.2\textwidth}{Teaching Social Studies} &      3.70 &      3.90 &      3.72 &      4.33 \\ 
 & (2.94)  & (3.15)  & (2.97)  & (3.13)  \\ 
\addlinespace[1ex] \multicolumn{5}{l}{\emph{Panel C. Student enrollment data}}  \\ 
\addlinespace[1ex] \multirow[t]{ 2}{0.2\textwidth}{Classrooms} &         . &         . &         . &     13.55 \\ 
 &  &  &  & (5.44)  \\ 
\addlinespace[1ex] \multirow[t]{ 2}{0.2\textwidth}{Pupils enrolled} &         . &         . &         . &   1056.49 \\ 
 &  &  &  & (   499.92)  \\ 
\addlinespace[1ex] \multirow[t]{ 2}{0.2\textwidth}{Pupil-teacher ratio} &         . &         . &         . &     51.30 \\ 
 &  &  &  & (    15.17)  \\ 
\addlinespace[1ex] Schools  & \multicolumn{1}{c}{      259}  & \multicolumn{1}{c}{      259}  & \multicolumn{1}{c}{      259}  & \multicolumn{1}{c}{      259}  \\ 
\addlinespace[1ex] 
\bottomrule
\end{tabular}

\end{center}
\vskip-2ex 
\floatfoot{
    \textsc{Notes}---Table presents school-level summary statistics, by data source.  Standard deviations in parentheses.   Sample consists of schools for which exam data were received from REB and that have corresponding entries in teacher placement data.  \emph{Average exam score} is the mean P6 exam score for students in that school and year; exam data are available for 2016--2018 only.  \emph{Number of teachers} is a count of all teaching staff.  For each subject, the table then reports the total number of teachers who teach at least one course in that subject. \emph{Classrooms} and \emph{Pupils enrolled} are provided in student enrollment data, which are available in 2019 only.  
}
\end{footnotesize}
\end{table}

Table \ref{t:schoolchars} summarizes exam performance and student enrollment in the matched sample of schools for which exam data are available, together with summaries of teacher deployment at the school level for the same schools.  P6 exam scores took an average value of between 34 and 37 points over these years.  These will be standardized within years to have a mean of zero and variance of one, with higher scores denoting better performance, in the analysis of Section \ref{s:results}. The schools in this sample had an average of between 20--22 employees over the period studied.  Patterns of subject coverage differed substantially across teachers, with around six teachers teaching each of the subjects of English, Kinyarwanda, and Math, but with a smaller number of teachers covering Science and Social Studies.  Finally, enrollment data show that sample schools had 13.6 classrooms and a total enrollment of 1056.5 on average in 2019. This implies a pupil-teacher ratio of 51.3 in sample schools. Because such enrollment figures are typically slow-moving, the analysis will treat pupil enrollment as a time-invariant school characteristic, which enables the use of pupil-teacher ratios as a predictor of teacher attrition in the analysis of Section \ref{s:results}.

\section{Results}\label{s:results}

In this section, I answer five questions.  First, I document the amount of teacher churn that exists.  Second, I examine the school-level correlates of teacher separations.  This speaks to how the burden of teacher turnover is shared across the school system.  Third, I examine what types of teachers are retained within schools.  Fourth, I speak to one measure of the human-resource challenge that this creates:  the challenge of sustaining stable school sizes in an environment in which teacher exits are common.  And fifth, I estimate the possible consequences of teacher loss for subsequent school performance.  

\subsection{How much churn is there?}\label{ss:incidence}

\begin{figure}[!hbtp]
\caption{Teacher retention outcomes}
\label{f:retention_rates}
\begin{footnotesize}

\begin{center}
\includegraphics[width=0.65\textwidth]{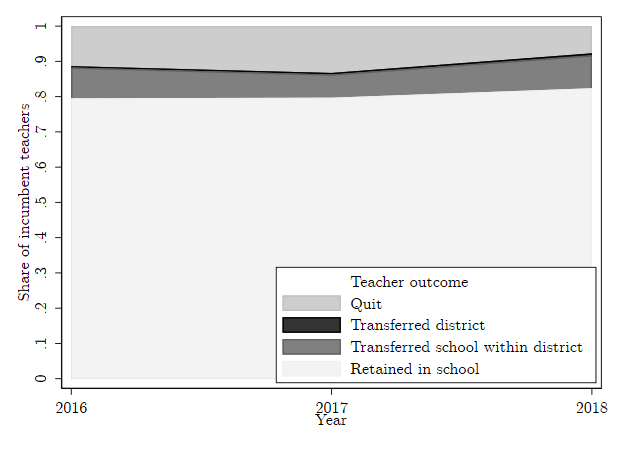}
\end{center}
    
\floatfoot{
    \textsc{Notes}---Figure shows subsequent retention outcomes of teachers employed in each year, as a proportion of total teaching.  Estimates based on sample of schools for which exam data are available.
}
\end{footnotesize}
\end{figure}

I estimate approximately 20 percent of teachers in any given year are no longer working in the same school in the subsequent year.  As illustrated in Figure \ref{f:retention_rates}, of those who separate from their jobs, approximately 11 percent quit teaching altogether; approximately 8 percent transfer to another school within the same district; and the remainder---less than 1 percent---transfer to another district.

These estimates are based on outcomes in the sample of schools for which I have exams data, because in that representative sample I also matched the names of schools as written in the teacher deployment records for consistency across years and across datasets. In the larger universe of administrative teacher placement data, school names are not consistently coded, though I do observe districts of placement and whether teachers are retained across years.  As Table \ref{t:TurnoverSummary} illustrates, those outcomes that \emph{are} observed in both the exam sample of schools and in the full set of teacher placements confirm the representativeness.  The rates of retention within districts in Panel B, which can be calculated by adding the fractions of teachers who remain in the school to the fraction of teachers who change schools within the district, are consistent across all years.

These estimates suggest a teacher attrition rate that is higher than typical of other educational systems.    Using nationally representative data from the United States, \citet{GolTaiRid14nces} estimate that approximately 8 percent of teachers exited annually in 2012-13, with a further 8 percent switching schools.    
Perhaps surprisingly, teacher exit rates among \emph{beginner} teachers were actually quite a bit lower in the U.S..  Of teachers beginning work in the 2007/08 school year, 48 percent were in the same school five years later; 13 percent were in another school in the same district; and 16 percent were in a different district, with the remaining 23 percent having left teaching for at least one year \citep{RauGra15nces}.  This five-year retention rate of 77 percent is consistent with a one-year retention rate of 94.9 percent. 
In a U.K. context, \citet{GibScrTel18cepr} find that 10.5 percent of teachers separate from their schools annually, with 7.8 percent of those separations representing moves to other schools, and 2.7 percent representing exits from the profession.  In Chile, \citet{BehTinTodWol16jole} low overall attrition rates---for example, fewer than 3 percent of teachers exit their municipal school system annually.  

\begin{table}[!hbtp]
\caption{Teacher turnover outcomes}
\label{t:TurnoverSummary}
\begin{footnotesize}
\begin{center}
\begin{tabular}{p{0.35\textwidth} *{3}{S}}
\toprule
 & \multicolumn{3}{c}{Year} \\ 
\cmidrule(lr){2-4}
\multicolumn{1}{c}{\text{ }} & \multicolumn{1}{c}{\text{2016}} & \multicolumn{1}{c}{\text{2017}} & \multicolumn{1}{c}{\text{2018}}\\
\midrule
\multicolumn{4}{l}{\emph{Panel A.  All teachers}}  \\ 
\addlinespace[1ex] Teacher remained in district  &      0.88 &      0.93 &      0.92 \\ 
\addlinespace[1ex] Teacher changed districts  &      0.00 &      0.00 &      0.01 \\ 
\addlinespace[1ex] Teacher left teaching profession  &      0.12 &      0.07 &      0.08 \\ 
\addlinespace[1ex] \multicolumn{4}{l}{\emph{Panel B.  Teachers in sample schools}}  \\ 
\addlinespace[1ex] Teacher remains in school  &      0.80 &      0.84 &      0.82 \\ 
\addlinespace[1ex] Teacher changed schools within district  &      0.09 &      0.07 &      0.09 \\ 
\addlinespace[1ex] Teacher changed districts  &      0.00 &      0.00 &      0.01 \\ 
\addlinespace[1ex] Teacher left teaching profession  &      0.11 &      0.09 &      0.08 \\ 
\addlinespace[1ex] \addlinespace[1ex] 
\bottomrule
\end{tabular}

\end{center}

\vskip-4ex
\floatfoot{
    \textsc{Notes}---Table summarizes subsequent placement outcomes for teachers present in schools in the year indicated in each column. Outcomes within each panel are mutually exclusive. School-level placement information is available for teachers in sample schools only; consequently, the \emph{Remained in district} category for the universe of teachers corresponds to the sum of the \emph{Remained in school} and \emph{Changed schools within districts} outcomes for teachers in the sample schools.
}

\end{footnotesize}
\end{table}

Less is documented about teacher turnover rates in low-income countries, and in sub-Saharan Africa in particular.  Heterogeneity in teachers' wage premia relative to outside options for similarly educated individuals \citep{EvYuaFil20cgd} suggests that teacher turnover is likely to vary across educational levels and between countries.  For Rwandan primary teachers, whose incomes are comparatively low, it appears that teacher turnover rates are substantially higher than those found in middle- and upper-income countries.

\subsection{What types of schools lose teachers?}\label{ss:school_attributes}

\begin{figure}[!hbtp]
\caption{Teacher school retention rates, by district}
\label{f:retention_map} 
\begin{center}
\includegraphics[width=0.65\textwidth]{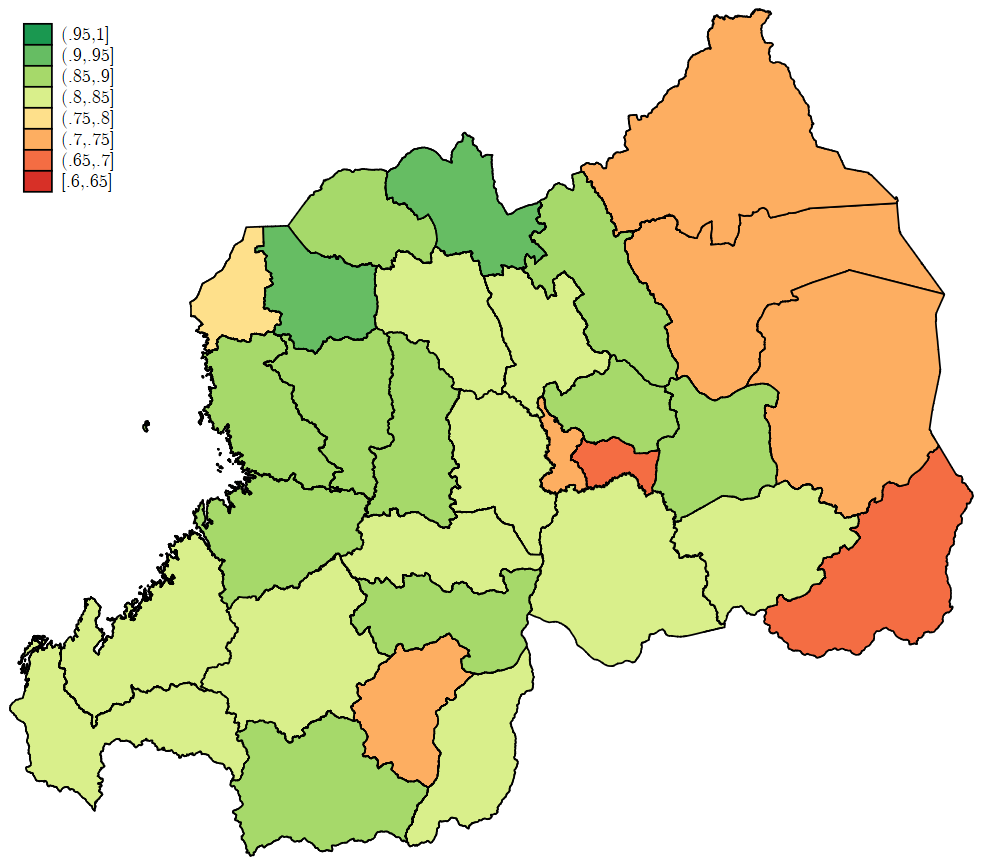}
\end{center}

\vskip-4ex 
\begin{footnotesize}
\floatfoot{
    \textsc{Notes}---Figure illustrates the share of teachers who remain in the same school in the subsequent year, across districts. Estimates based on sample of schools for which exams data are available. Estimates are average retention rates among teachers employed in each of the years 2016--2018; retention outcomes for two districts (Gasabo and Nyarugenge) are not available for 2017.
}
\end{footnotesize}
\end{figure}

A first indication of the incidence of the churn problem across schools can be obtained from the geographic distribution of teacher exits.  This is illustrated in Figure \ref{f:retention_map}.  There, I highlight the share of teachers who remain in the same school across years.  These school retention rates are averaged at the district level, combining information from the three years for which retention outcomes are observed in sample schools.

Figure \ref{f:retention_map} makes clear that the challenge of teacher retention is not confined to either rural or urban areas.\footnote{Exact figures for each district are provided in Appendix Table \ref{t:retention_by_district_year}.}  Rather, there are pockets of challenge in several regions of the country.  The two lowest retention rates occur in Kicukiro (66 percent) and Kirehe (68 percent), suggesting that cost of living in Kigali districts may feature alongside other aspects of teacher employment conditions and labor supply.  Zooming out, there is a swath of districts with retention rates below 75 percent in the Eastern Province, as well as Huye District in the Southern Province. 

Beneath this geographic variation, the incidence of teacher churn varies by type of school.  To demonstrate this pattern of incidence, I estimate a model that explains the share of teachers that exit following year $t$ in school $s$ of district $d$ as a function of school characteristics, $x_{sdt}$, and district-year fixed effects, $\mu_{dt}$:
    \begin{equation}\label{eq:incidence-schools}
    \text{Share lost}_{sdt} = \beta x_{sdt} + \mu_{dt} + e_{sdt}
    \end{equation}

I focus on three key characteristics of interest:  exam performance, pupil-teacher ratios, and school size.  

\begin{table}[!h]\caption{Share of teachers leaving the school, by school-year attributes}
\begin{footnotesize}
\begin{center}
\begin{tabular}{l *{4}{S}}
\toprule
\multicolumn{1}{c}{\text{ }} & \multicolumn{1}{c}{\text{(1)}} & \multicolumn{1}{c}{\text{(2)}} & \multicolumn{1}{c}{\text{(3)}} & \multicolumn{1}{c}{\text{(4)}}\\
\midrule
Exam score  &     -0.01\ensuremath{^{*}} &         . &         . &     -0.01 \\ 
 & (0.01)  &  &  & (0.01)  \\ 
\addlinespace[1ex] Ln pupil-teacher ratio  &         . &     -0.10\ensuremath{^{***}} &         . &     -0.09\ensuremath{^{***}} \\ 
 &  & (0.02)  &  & (0.02)  \\ 
\addlinespace[1ex] Ln enrollment  &         . &         . &     -0.02 &     -0.00 \\ 
 &  &  & (0.01)  & (0.01)  \\ 
\addlinespace[1ex] Observations  & \multicolumn{1}{c}{      741}  & \multicolumn{1}{c}{      726}  & \multicolumn{1}{c}{      726}  & \multicolumn{1}{c}{      726}  \\ 
\addlinespace[1ex] 
\bottomrule
\end{tabular}

\end{center}

\vskip-4ex
\floatfoot{
    \textsc{Notes}---Dependent variable is share of teachers separating from the school following the school year in question. \emph{Exam score} is the (standardized) average P6 exam result; \emph{pupil-teacher ratio} and \emph{enrollment} are included in logs. Enrollment data are 2019 values. All specifications include district-year fixed effects; sample is restricted to school-years for which exam score outcomes are known.     
}
\end{footnotesize}
\end{table}

Better performing schools lose fewer teachers.  A one standard-deviation increase in exam scores is associated with a reduction of one percentage point in the share of teachers leaving the school after that year. Of course, this association may reflect causality in either direction.  Serially correlated teacher exits may cause declines in exam scores; the potential educational consequences of teacher exits are discussed further in Section \ref{ss:learning}. Teachers may find schools with attributes that cause poor learning less appealing---or may be deterred by learning failures directly.  Regardless of the causal story that underpins this association, it suggests that the challenge of teacher churn threatens to exacerbate educational inequalities.

Surprisingly, the relationship between the pupil-teacher ratio and the share leaving is negative. As column (2) shows,  schools with high pupil-teacher ratios are actually \emph{less} likely to lose additional teachers.  This does not appear to be explicable entirely in terms of urban schools, which have higher enrollment levels, retaining more teachers. Although statistically indistinguishable from zero, the relatively precise school-size association estimated in column (3) allows us to rule out positive associations between enrollments and turnover of any substantial magnitude.  And in column (4), I see that the observed relationship between pupil-teacher ratios and subsequent teacher separations survives inclusion of controls for exam performance and pupil enrollment. The negative relationship between PTR and teacher turnover persists even when Equation \eqref{eq:incidence-schools} is estimated with school fixed effects; it is also robust to the inclusion of prior lags of or changes in PTR, as well as a transformation of the outcome to be an indicator variable for whether the school loses \emph{any} teachers (not shown).  While associational, this result challenges a simple narrative that teacher workloads, in the form of large class sizes, drive turnover.  Alternatively, it may be the case that both pupils and teachers respond positively to the same signals of school quality improvements, sorting positively into the same, improving schools over time.

\subsection{What types of teachers leave?}\label{ss:teacher_attributes}

Not all forms of teacher turnover are equal:  the associated losses (or gains) to the productivity of the education sector will depend on the skill and experience profile of teachers who leave.  In this section, I explore how teacher characteristics are associated with the likelihood that they separate from a specific school, and that they exit the teaching sector altogether.

To do so, I use data from the matched sample of schools to estimate a linear probability model of the form
\begin{equation}\label{eq:teacherhazard}
\Pr(y_{it}=1|x_i,\text{tenure}_{it},d,t) = \beta x_i + f(\text{tenure}_{it}) + \delta_d + \gamma_t,
\end{equation}
where $x_{i}$ are attributes of teacher $i$, $\text{tenure}_{it}$ is teacher $i$'s tenure in year $t$, and $d$ and $t$ denote the district and year associated with the teacher. This model allows the baseline probability of a teacher's exit to be a flexible function, $f(\cdot)$, of their years of tenure.  I estimate Equation \eqref{eq:teacherhazard} both to predict the likelihood that a teacher separates from their job in a particular school---outcome \emph{Separation} in Models A and B---and that a teacher quits public-sector teaching---outcome \emph{Quit} in Models C and D. In Models A and C, I allow only the fixed attributes of teachers to serve as predictors of their probability of attrition.  In Models B and D, I allow the corresponding forms of attrition to further depend on the nature of the teacher's assignments within the school, allowing us to understand how the market return to different skills affects subject-specific shortages of staff.   

\begin{table}[hbtp]
\caption{Hazard of teacher exit} 
\label{t:QuitHazard}
\begin{footnotesize}
    \begin{center}
        \begin{tabular}{p{0.2\textwidth} *{4}{S}}
\toprule
 & \multicolumn{2}{c}{Separation} & \multicolumn{2}{c}{Quit}  \\ 
\cmidrule(lr){2-3} \cmidrule(lr){4-5} 
\multicolumn{1}{c}{\text{ }} & \multicolumn{1}{c}{\text{Model A}} & \multicolumn{1}{c}{\text{Model B}} & \multicolumn{1}{c}{\text{Model C}} & \multicolumn{1}{c}{\text{Model D}}\\
\midrule
Female  &     -0.01\ensuremath{^{**}} &     -0.01\ensuremath{^{*}} &     -0.01\ensuremath{^{*}} &     -0.01\ensuremath{^{*}} \\ 
 & (0.01)  & (0.01)  & (0.01)  & (0.01)  \\ 
\addlinespace[1ex] Tenure $\leq$ 0  &      0.39\ensuremath{^{***}} &      0.39\ensuremath{^{***}} &      0.32\ensuremath{^{***}} &      0.32\ensuremath{^{***}} \\ 
 & (0.03)  & (0.03)  & (0.03)  & (0.03)  \\ 
\addlinespace[1ex] Tenure $\leq$ 1  &      0.16\ensuremath{^{***}} &      0.16\ensuremath{^{***}} &      0.03\ensuremath{^{*}} &      0.03\ensuremath{^{*}} \\ 
 & (0.06)  & (0.06)  & (0.01)  & (0.01)  \\ 
\addlinespace[1ex] Tenure $\leq$ 2  &      0.07\ensuremath{^{***}} &      0.06\ensuremath{^{***}} &      0.01 &      0.01 \\ 
 & (0.02)  & (0.02)  & (0.02)  & (0.02)  \\ 
\addlinespace[1ex] Tenure $\leq$ 3  &      0.07\ensuremath{^{***}} &      0.07\ensuremath{^{***}} &      0.01 &      0.01 \\ 
 & (0.02)  & (0.02)  & (0.02)  & (0.02)  \\ 
\addlinespace[1ex] Tenure $\leq$ 4  &      0.05\ensuremath{^{***}} &      0.05\ensuremath{^{***}} &      0.02 &      0.02 \\ 
 & (0.02)  & (0.02)  & (0.01)  & (0.01)  \\ 
\addlinespace[1ex] Tenure $\leq$ 5  &      0.05\ensuremath{^{***}} &      0.05\ensuremath{^{***}} &      0.01 &      0.01 \\ 
 & (0.02)  & (0.02)  & (0.01)  & (0.01)  \\ 
\addlinespace[1ex] Tenure $\leq$ 7.5  &      0.01 &      0.01 &     -0.01 &     -0.01 \\ 
 & (0.02)  & (0.02)  & (0.01)  & (0.01)  \\ 
\addlinespace[1ex] Tenure $\leq$ 10  &      0.01 &      0.00 &     -0.01 &     -0.01 \\ 
 & (0.02)  & (0.02)  & (0.01)  & (0.01)  \\ 
\addlinespace[1ex] Tenure $\leq$ 12.5  &      0.00 &      0.01 &     -0.01 &     -0.01 \\ 
 & (0.02)  & (0.02)  & (0.01)  & (0.01)  \\ 
\addlinespace[1ex] Tenure $\leq$ 15  &      0.01 &      0.01 &     -0.02 &     -0.02 \\ 
 & (0.02)  & (0.02)  & (0.01)  & (0.01)  \\ 
\addlinespace[1ex] Tenure $\leq$ 17.5  &     -0.02 &     -0.02 &     -0.02 &     -0.02 \\ 
 & (0.02)  & (0.02)  & (0.01)  & (0.01)  \\ 
\addlinespace[1ex] Tenure $\leq$ 20  &     -0.03\ensuremath{^{*}} &     -0.02 &     -0.04\ensuremath{^{***}} &     -0.03\ensuremath{^{***}} \\ 
 & (0.02)  & (0.02)  & (0.01)  & (0.01)  \\ 
\addlinespace[1ex] Tenure $\leq$ 22.5  &     -0.03 &     -0.02 &     -0.01 &     -0.01 \\ 
 & (0.02)  & (0.02)  & (0.02)  & (0.02)  \\ 
\addlinespace[1ex] Tenure $\leq$ 25  &     -0.00 &      0.00 &     -0.02 &     -0.02 \\ 
 & (0.02)  & (0.02)  & (0.02)  & (0.02)  \\ 
\addlinespace[1ex] Teaching English  &         . &      0.01 &         . &      0.01 \\ 
 &  & (0.01)  &  & (0.01)  \\ 
\addlinespace[1ex] Teaching Kinyarwanda  &         . &     -0.01 &         . &     -0.00 \\ 
 &  & (0.01)  &  & (0.01)  \\ 
\addlinespace[1ex] Teaching Math  &         . &      0.01 &         . &      0.01\ensuremath{^{**}} \\ 
 &  & (0.01)  &  & (0.01)  \\ 
\addlinespace[1ex] Teaching Science  &         . &      0.01 &         . &      0.00 \\ 
 &  & (0.01)  &  & (0.01)  \\ 
\addlinespace[1ex] Teaching Social Studies  &         . &      0.00 &         . &     -0.00 \\ 
 &  & (0.01)  &  & (0.01)  \\ 
\addlinespace[1ex] Observations  & \multicolumn{1}{c}{15807}  & \multicolumn{1}{c}{15435}  & \multicolumn{1}{c}{15807}  & \multicolumn{1}{c}{15435}  \\ 
\addlinespace[1ex] 
\bottomrule
\end{tabular}

    \end{center}

\vskip-7ex 

\floatfoot{
    
    \textsc{Notes}---Table contains estimates from alternative proportional hazard models, estimated via logistic regression.  Outcome in Models A and B is an indicator for whether the teacher in question separated from their job in a given school in that year; outcome in Models C and D is an indicator for whether the teacher quit the teaching profession in that year.  Variables English, Kinyarwanda, Math, Science, and Social Studies are indicators for whether the teacher taught that subject in the prior year.  Omitted category for tenure-specific hazard rates are teachers whose tenure exceeds 20 years. All estimates include District and Year indicators (not shown).

}
\end{footnotesize}
\end{table}

\begin{figure}[!ht]
\caption{Expected probabilities of teacher separations and quits, by tenure}
\label{f:hazards}
\begin{footnotesize}
\begin{center}
\includegraphics[width=0.65\textwidth]{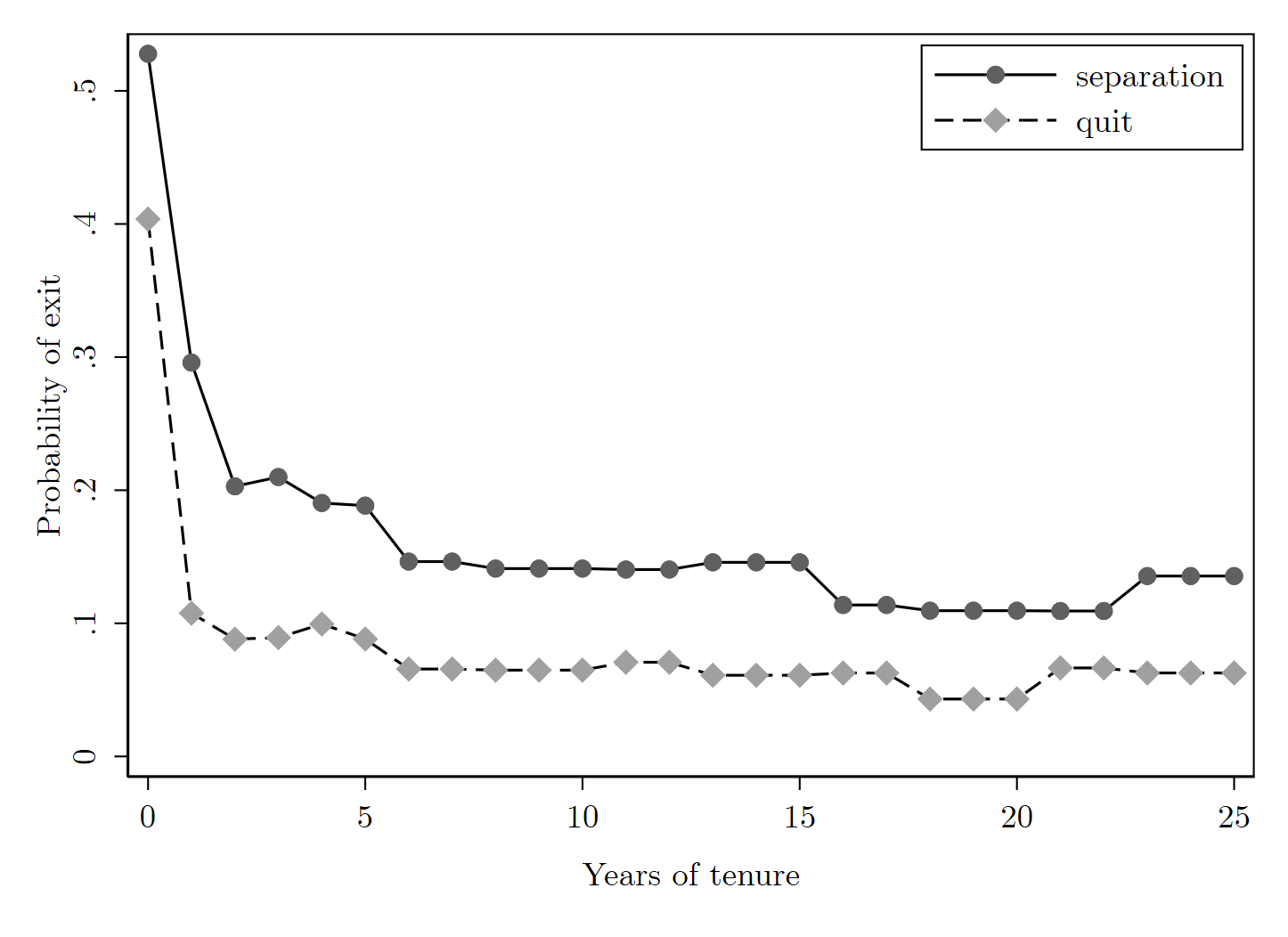}
\end{center}

\vskip-6ex 

\floatfoot{
    \textsc{Notes}---Figure displays the probability that a teacher with $t$ years of tenure separates from their job, or quits the teaching workforce, respectively.  Estimates derived from Models A and C of Table \ref{t:QuitHazard}.   
}
\end{footnotesize}
\end{figure}

Results from Models A and C show dramatic differences in the likelihood of both job separations and quits across the tenure profile of teachers.  This is illustrated in Figure \ref{f:hazards}, which shows the annual hazard rate for teacher job separations in general, and for teacher exits from the public-sector workforce in particular. These estimates suggest dramatic exit rates for early-career teachers:  40 percent of first-year teachers exit the work force each year, and annual quit rates remain above 10 percent through the fifth year of teacher employment.  The gap between quits and separations reflects both within-district and across-district transfers, and appears to narrow gradually over the course of teachers' employment histories.  Taken together, these results suggest that job loss is greatest among early-career teachers, and that addressing this challenge by ensuring that the best early-stage teachers are retained will be an important part of any strategy to expand the teaching workforce. 

Women are estimated to be statistically significantly less likely to separate from their jobs or to quit the education sector, across all models.  In each year, their likelihood of departure is less by 1 percentage point.  These differences compound across years, implying a growing representation of women in the gender composition of teachers across the tenure profile.

Among subjects taught, teaching math is associated with a statistically significant, 1.3 percentage point higher likelihood of quitting.  This is  a substantial difference when compared with the average quit rate of between 8 and 12 percent per year estimated in Table \ref{t:TurnoverSummary}. This is consistent with the presence of strong outside opportunities for individuals with the skills required to teach Mathematics effectively. 

Taken together, these findings are consistent with a view that losses are concentrated among those whose outside opportunities in the labor market are comparatively strong. These include new teachers, who by virtue of their educational experiences may be particularly likely to be proficient in English, and teachers with quantitative skills, proxied by assignment to mathematics.  In other contexts, teachers whose value added is greater have been found to have better-paying outside options; see \citet{FigKen91jpubec} for evidence from the U.S., and \citet{TeaXXjae} for a discussion of the implications of such sorting in sub-Saharan African labor markets more generally.  To the extent that this is also true in Rwanda, it heightens the importance of targeted retention policies that improve the attraction of the teaching profession specifically for those teachers with the skills to make the greatest contribution to student learning.

\subsection{Where is replacement hard?}\label{ss:replacement}

If school districts have difficulty replacing those teachers who leave, teacher turnover can create distinct and direct challenges for student learning.  The resulting short-term staffing shortages in schools not only stretch pupil-teacher ratios, but may also require schools to assign teachers to teach subjects outside their areas of expertise.  This can lower teachers' ability to impart effective instruction in ways that ways that are not visible in the raw pupil-teacher ratio.  

One measure of the efficiency with which district human resource systems function is the likelihood that a given school, in the year after losing one or more teachers, is able to acquire new staff members to avoid decreasing in total staff size. This measure will capture the specific targeting of new teachers to schools that have lost staff, though it may be easier for districts doing large-scale hiring to perform well on such a metric. Since in our matched sample I observe more than 600 school-years in which a school loses at least one teacher, I can estimate this probability of complete replacement on average for each of the districts.

\begin{figure}[!hbtp]
\caption{Replacement rates:  Probability that a school losing one or more teachers is able to maintain its size in the subsequent year}
\label{f:replacement}

\begin{footnotesize}
\begin{center}
\includegraphics[width=0.65\textwidth]{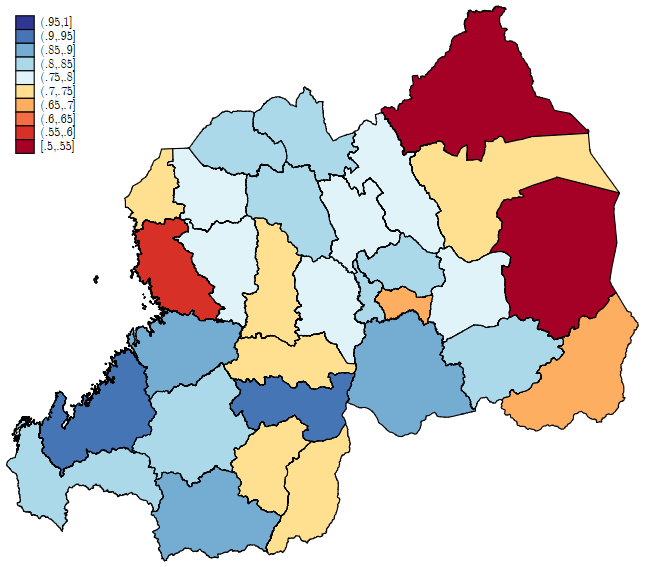}
\end{center}

\vskip-4ex

\floatfoot{
    \textsc{Notes}---Figure estimates the probability in each district that a school retained or increased its size, conditional on losing at least one teacher in the prior year. 
}
\end{footnotesize}
\end{figure}

In the year following a job separation, districts were able to keep schools at or above their prior staff size 77 percent of the time on average. Figure \ref{f:replacement} visualizes the variation in this replacement frequency across districts.  There appears to be quite substantial variation in these outcomes across districts. In Kayonza, Nyagatare, and Rutsiro, fewer than 60 percent of schools that lose teachers are able to maintain their size in the subsequent year.  At the high end, in Nyamasheke and Nyanza, more than 90 percent of matched-sample schools are able to retain their size after losing a teacher.  

This heterogeneity in replacement rates across districts may reflect a mixture of demand- and supply-side conditions. On the demand side, districts that receive information quickly, and respond quickly to advertise and recruit for vacancies will tend to fill those vacancies more quickly.   On the labor-supply side, districts with a sizeable stock of individuals qualified to teach and with amenities favorable to selection into teaching may all see greater interest in vacancies and faster replacement rates. Kigali districts are not necessarily those with the highest rates, suggesting that cost of living and the strength of outside opportunities may be important challenges to high replacement rates.

\subsection{What are the learning consequences of teacher departures?} \label{ss:learning}

The policy importance of teacher turnover arises primarily due to the threat that it may have adverse consequences for learning outcomes.  Section \ref{ss:replacement} has shown that the replacement of teachers who leave schools can take time, and that the ability of districts to do so is uneven.  Resulting shocks to pupil-teacher ratios may have a direct effect on learning outcomes. 

Learning may also be adversely affected to the extent that schools coping with staffing shortages are forced to have teachers teach in subjects outside of their areas of qualification.  The prevalence of \emph{cross-specialty teaching} provides a lower bound on the prevalence of this practice.  Since teachers are certified in one of three subject specialties, when teaching assignment data show them teaching in multiple subjects that span these Teacher Training Certificate subject families, they must be teaching in at least one subject outside their area of specialty.\footnote{Specifically, these Teacher Training Certificate specialties are \emph{Modern Languages} (English and Kinyarwanda), \emph{Math and Science}, and \emph{Social Studies}.}  Between 2016--2019, the data contain 148,987 teacher-years in which subjects taught are reported, allowing us to examine the prevalence of cross-specialty teaching.

On average across these four years, 21.7 percent of teachers are teaching in subjects that cross borders of specialization. This represents a lower bound on the prevalence of out-of-specialty teaching among primary school teachers, since a teacher may be teaching \emph{entirely} out of their subject in a given year.  Combining information on a given teacher from across all four years, the majority of teachers---fully 50.6 percent of the 55,555 teachers whose subjects taught are reported in at least one year over this period---have taught in two or more subjects belonging to distinct Teacher Training Certificate families.   

Motivated by this evidence, I use the panel structure of the matched sample of schools to estimate the consequences of teacher exits on subsequent student learning outcomes.  These estimates should be taken with a great deal of caution, since teacher separations are not randomly assigned.  But given the potential magnitude of this challenge, I put forward a set of observational estimates that I take to be indicative of the possible range of causal effects.

To do so, I estimate the effects of teacher separations on subsequent school learning outcomes with a model of the following form:
\begin{equation}
\label{eq:impacts_separation}
y_{sd,t+1} = \tau \cdot \text{Separation}_{isdt} + f(y_{sdt}) + \beta x_{isdt}  + \mu_{d,t+1} + e_{sd,t+1}.
\end{equation} 
Here, $y_{sd,t+1}$ represents average test scores in school $s$ of district $d$ in year $t+1$; these are standardized to have a mean of zero and standard deviation of one within each year, with higher values denoting better outcomes.  Our primary interest is in the parameter $\tau$, which estimates the effect of teachers $i$ separating from the school between years $t$ and $t+1$ on exam scores in year $t+1$. I allow outcomes in year $t$ to depend on a cubic function of prior-year test scores, $f(y_{sdt})$.  I also control for a range of characteristics, $x_{isdt}$, of both the teacher and the school that may be correlated with both teacher separations and subsequent exam performance.  

\begin{table}[!hbtp]
\caption{Exam score changes associated with teacher retention}
\label{t:LossEffects}
\begin{footnotesize}
\begin{center}
\begin{tabular}{l *{4}{S}}
\toprule
\multicolumn{1}{c}{\text{ }} & \multicolumn{1}{c}{\text{(1)}} & \multicolumn{1}{c}{\text{(2)}} & \multicolumn{1}{c}{\text{(3)}}\\
\midrule
Teacher separation  &     -0.05\ensuremath{^{*}} &     -0.05\ensuremath{^{*}} &     -0.09\ensuremath{^{**}} \\ 
 & (0.03)  & (0.03)  & (0.04)  \\ 
\addlinespace[1ex] Observations  & \multicolumn{1}{c}{    10018}  & \multicolumn{1}{c}{    10018}  & \multicolumn{1}{c}{     4704}  \\ 
\addlinespace[1ex] 
\bottomrule
\end{tabular}

\end{center}

\vskip-4ex 

\floatfoot{
    \textsc{Notes}---Table presents estimates of the impact of teacher exit on subsequent year school exam-score performance. Unit of observation is the teacher-year, with standard errors clustered at the school level. All estimates control for district-year fixed effects, for a cubic function of lagged test scores, for log total teachers in the school and for log total pupils enrolled. Model (1) includes indicators for each of three tenure categories (0--5 years, 6--15 years, 16+ years), for subjects to which the teacher is assigned, and for teacher gender.  Model (2) controls for the school's overall workforce composition by adding (IHS-transformed) sums of the count of teachers in each of the categories in Model (1), at the school level.  Model (3) includes two-year lags of cubic polynomial in exam scores.   
}
\end{footnotesize}
\end{table}

In our baseline specification, controls include log total enrollment and log total numbers of teachers, as well as indicators for teacher gender, for each of three levels of tenure (less than 5 years, 6--15 years, and 16 or more years, each of which comprises approximately a third of the sample), and for core-curricular subjects taught.  Our second model adds (inverse-hyperbolic-sine transformed) aggregates of these teacher-level characteristics at the school level, to control for the overall staffing configuration of the school.  And to address the possibility that schools losing teachers may differ not only in terms of levels but also in the \emph{trajectories} of their learning outcomes, our third model adds a flexible, cubic function of lagged test scores alongside the test scores of year $t$.
While this is necessarily an observational analysis, and the assumption that the association between time-varying unobserved confounders and teacher losses is strong, this analysis nonetheless provides an indication of the potential consequences of teacher loss, and can help inform thinking about the magnitude of omitted variable bias required to overturn the substantive implications of these findings.

The average exam-score gains associated with losing a teacher are reported in Table \ref{t:LossEffects}. 
Estimates are remarkably consistent across each of these models, suggesting that teacher exits are associated with declines in subsequent-year test scores of between 0.05 and 0.09 standard deviations.  Pooled estimates of these effects on outcomes in 2017 and 2018 exams in columns (1) and (2)---the latter of which adds a rich array of controls for prior staffing levels---are statistically significant at the 10 percent level.   Moreover, this finding appears not to be driven by differential learning trends in schools that lose teachers. Model (3) tests for such differential trends by controlling for two years of lagged test scores.  Point estimates in this model are actually larger in magnitude, suggesting a loss of 0.09 standard deviations for each teacher exit.  That finding is significant at the 5 percent level, in spite of the reduced sample size that arises from the fact that two-year learning histories are available for schools only in 2018.  This finding suggests that this form of omitted variable bias is, if anything, causing our pooled model to \emph{underestimate} the test-score benefits of teacher retention.  

If these observational estimates are taken as causal, such effect sizes appear to be substantial in policy terms.  Conceptually, it is not straightforward to compare the consequences of teacher losses to experimental or quasi-experimental estimates of the effect sizes of educational interventions, the latter of which should be thought of in relation to their costs \citep{Kra20edresearcher}.\footnote{A further issue is that if outcomes are compressed due to low overall performance, a one-standard deviation difference in learning outcomes may mean less than it does in a setting with more meaningful variation in test scores. In the absence of an international assessment to provide a cross-walk to other countries, it is not possible to translate learning differences from Rwanda onto the same \emph{absolute} scale as estimates from other countries. However, the facts that estimates here are for summative, Primary 6 exams, and that they show no evidence of floor effects (Appendix Figure \ref{f:exam_histogram}), suggests that observed effects sizes are consequential.}  Nonetheless, a recent review of effect sizes in studies of education interventions in low- and middle-income countries finds median effect sizes of 0.10 standard deviations \citep{EvaYua20cgd}, only slightly larger in magnitude than the learning losses associated with a single teacher job separation.  The magnitude of learning losses that follow a teacher separation are surprising because the teachers lost may not have direct teaching responsibility for the exam-taking year; if causal, these estimates would represent an average of the direct consequences of losing a Primary 6 teacher and the indirect consequences of staffing shortages induced by teacher exits at other grade levels. 

\begin{figure}[!th]
\caption{Exam score changes associated with losing one or more teachers, by attributes of retained teachers}
\label{f:LossEffects}
\begin{footnotesize}
\begin{center}
\includegraphics[width=0.95\textwidth]{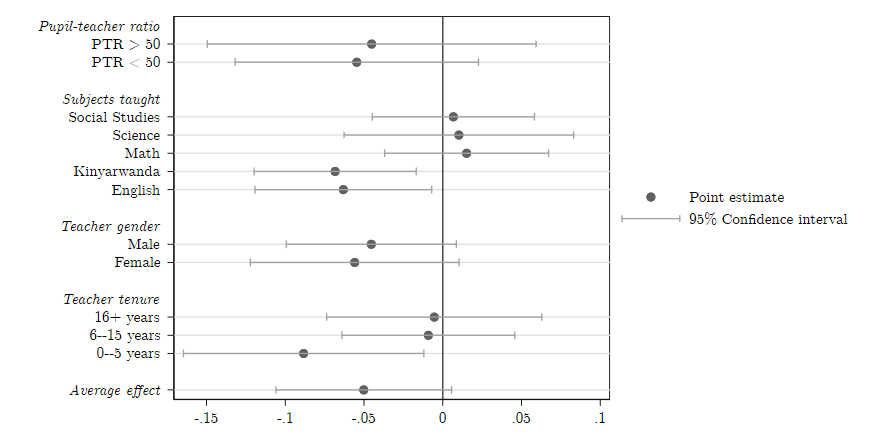}
\end{center}

\floatfoot{
    \textsc{Notes}---Figure illustrates point estimates and 95 percent confidence intervals for subgroup-specific estimates of changes in subsequent exam scores following the loss of at least one teacher.  Outcome is school-average Primary 6 exam score, pooling outcomes for 2017 and 2018. All specifications include controls for log enrollment and log numbers of teachers, as well as controls for teacher attributes and subject-specific staffing aggregates in the year prior to the exam.  Standard errors clustered at school level.
}

\end{footnotesize}
\end{figure}

Retention policies can be designed to prioritize retention of those teachers whose loss would have the greatest consequences for subsequent learning.  To illustrate the basis for such policies, in Figure \ref{f:LossEffects} I adapt the specification of Table \ref{t:LossEffects}, Model (2).  There, I visualize estimates that allow the impacts of teacher separations to vary by their type, along four dimensions:  subjects taught, tenure, gender, and school pupil-teacher ratios.  This is done by replacing the indicator for teacher separation with an interaction between teacher separation and indicators for teacher types, taking each dimension of teacher attributes as a separate regression.  

Figure \ref{f:LossEffects} highlights that the losses associated with losing early-career teachers are greatest---precisely those teachers for whom exit rates are highest. I find no meaningful difference between the effects of losing male and female teachers.  And among subjects taught, the loss of a teacher who teaches English or Kinyarwanda are greatest.  Finally, dividing schools into those with pupil-teacher ratios above and below 50 (the median is 54 in this sample), I find no meaningful difference in the effects of teacher loss between high-PTR and low-PTR schools.

\section{Conclusions}\label{s:conclusions}

This paper has documented the incidence of teacher turnover in Rwanda.  Levels of job separations have been found to be high, disproportionately borne by low-performing schools, and consequential for staffing continuity and learning outcomes.  Moreover, the pattern of attrition across teachers exacerbates the human resource challenge:  attrition rates are particularly acute among teachers in the first five years of their career, and the loss of exactly those teachers is associated with the greatest harms to student learning. 

I conclude by discussing three potential implications of these findings for policy. 

First, education systems should track teachers' movements as a vital statistic.  Typical systems focus on  \emph{static} snapshots, such as pupil-teacher ratios or physical and monetary resources per student. While certainly important, this focus has drawn attention away from the churn that lies below the surface, and that exacts an independent toll on learning outcomes.    

Second, the design of human-resource policies should consider the type of teacher turnover that they induce as a key intermediate policy outcome.  Policies regarding transfers across schools may be adapted to provide incentives for continuity in schools that need it.  This paper finds particularly high teacher turnover rates among early-career teachers, and finds the consequences of these separations to be disproportionately large.  Understanding how to account for the preferences of teachers in these initial placements, and how to retain the most effective teachers, may both be part of a strategy to mitigate the harms of these separations and improve the overall stock of teachers.  In Rwanda, the government's recent reforms of teacher recruitment and deployment provide an opportunity to design and test policies with these aims in mind.\label{text:OngoingReforms}  

Third, the efficiency with which human resource systems can deliver replacements for teachers lost from specific schools may be an avenue to improve learning outcomes.  This can reduce the extent to which schools have to adapt to transitory staffing shortages. Reliance on coping strategies for transitory staffing shortages, such as having teachers teach outside their subjects of expertise, may result in poor learning outcomes.  Efficient replacement of lost teachers can help to ensure that the staffing resources of schools remain aligned with their needs, for a given rate of teacher separations.

Taking these findings together, I suggest that teacher turnover is an under-appreciated challenge for developing-country school systems in countries like Rwanda.  The optimal rate of teacher turnover is certainly not zero.  Some turnover is inevitable.  It is appropriate to focus on the retention of those teachers who deliver the best learning outcomes, and those who teach in under-performing schools.  In spite of the substantial challenge that this phenomenon presents, there is considerable opportunity for policy to mitigate harms and improve learning outcomes.

\clearpage
\bibliographystyle{aer}
\bibliography{REB}

\cleardoublepage
\appendix
\noappendicestocpagenum \addappheadtotoc
\makeatletter
\def\@seccntformat#1{Appendix\ \csname the#1\endcsname\quad}
\def\@subseccntformat#1{\csname the#1\endcsname\quad}
\makeatother
\renewcommand\thetable{\Alph{section}.\arabic{table}}
\renewcommand\thefigure{\Alph{section}.\arabic{figure}}
\counterwithin{figure}{section}
\counterwithin{table}{section}

\clearpage
\section{Supplementary figures}
\label{s:supplementary_figures}

\begin{figure}[!h]
\caption{Distribution of school-average exam scores, 2016--2018}
\label{f:exam_histogram}
	\begin{footnotesize}
	\begin{center}
	\includegraphics[width=0.65\textwidth]{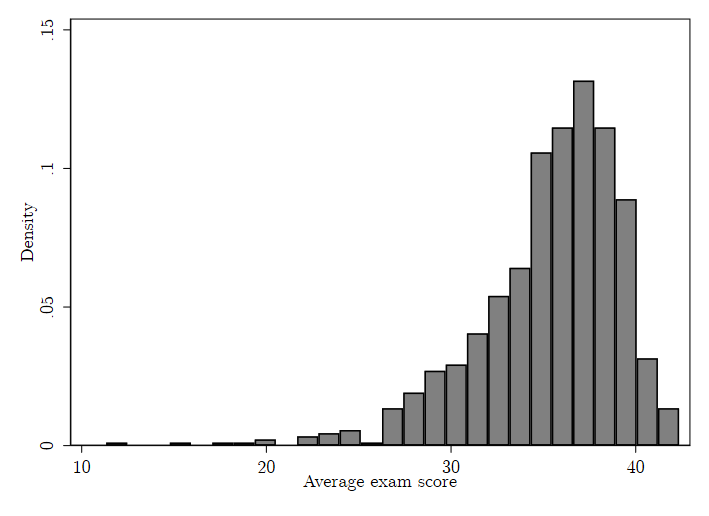}
	\end{center}

	\floatfoot{
	     \textsc{Notes}---Figure shows the distribution of school-level averages of Primary 6 exam scores in study sample schools, over the period from 2016--2018.
	}
	\end{footnotesize}
\end{figure}

\clearpage
\section{Supplementary tables}
\label{s:supplementary_tables}

\begin{table}[!h]
\caption{School retention rates by district and year}
\label{t:retention_by_district_year}

	\begin{footnotesize}
	    \begin{center}
		\resizebox{.5\linewidth}{!}{
	    \begin{tabular}{p{0.3\textwidth} *{3}{S}}
\toprule
 & \multicolumn{3}{c}{Year} \\ 
\cmidrule(lr){2-4}
\multicolumn{1}{c}{\text{District}} & \multicolumn{1}{c}{\text{2016}} & \multicolumn{1}{c}{\text{2017}} & \multicolumn{1}{c}{\text{2018}}\\
\midrule
\multicolumn{4}{l}{\emph{Kigali City}}  \\ 
\addlinespace[1ex] Gasabo  &      0.86 &         . &         . \\ 
\addlinespace[1ex] Kicukiro  &      0.82 &      0.46 &      0.72 \\ 
\addlinespace[1ex] Nyarugenge  &      0.73 &         . &         . \\ 
\addlinespace[1ex] \multicolumn{4}{l}{\emph{Eastern Province}}  \\ 
\addlinespace[1ex] Bugesera  &      0.88 &      0.77 &      0.84 \\ 
\addlinespace[1ex] Gatsibo  &      0.62 &      0.90 &      0.75 \\ 
\addlinespace[1ex] Kayonza  &      0.68 &      0.78 &      0.66 \\ 
\addlinespace[1ex] Kirehe  &      0.78 &      0.67 &      0.62 \\ 
\addlinespace[1ex] Ngoma  &      0.82 &      0.80 &      0.88 \\ 
\addlinespace[1ex] Nyagatare  &      0.70 &      0.63 &      0.82 \\ 
\addlinespace[1ex] Rwamagana  &      0.81 &      0.87 &      0.89 \\ 
\addlinespace[1ex] \multicolumn{4}{l}{\emph{Northern Province}}  \\ 
\addlinespace[1ex] Burera  &      0.89 &      0.95 &      0.89 \\ 
\addlinespace[1ex] Gakenke  &      0.89 &      0.90 &      0.71 \\ 
\addlinespace[1ex] Gicumbi  &      0.86 &      0.89 &      0.84 \\ 
\addlinespace[1ex] Musanze  &      0.87 &      0.89 &      0.83 \\ 
\addlinespace[1ex] Rulindo  &      0.75 &      0.89 &      0.90 \\ 
\addlinespace[1ex] \multicolumn{4}{l}{\emph{Southern Province}}  \\ 
\addlinespace[1ex] Gisagara  &      0.85 &      0.80 &      0.77 \\ 
\addlinespace[1ex] Huye  &      0.70 &      0.78 &      0.74 \\ 
\addlinespace[1ex] Kamonyi  &      0.86 &      0.89 &      0.75 \\ 
\addlinespace[1ex] Muhanga  &      0.83 &      0.91 &      0.85 \\ 
\addlinespace[1ex] Nyamagabe  &      0.86 &      0.78 &      0.84 \\ 
\addlinespace[1ex] Nyanza  &      0.92 &      0.88 &      0.81 \\ 
\addlinespace[1ex] Nyaruguru  &      0.82 &      0.90 &      0.89 \\ 
\addlinespace[1ex] Ruhango  &      0.69 &      0.89 &      0.93 \\ 
\addlinespace[1ex] \multicolumn{4}{l}{\emph{Western Province}}  \\ 
\addlinespace[1ex] Karongi  &      0.87 &      0.89 &      0.89 \\ 
\addlinespace[1ex] Ngororero  &      0.92 &      0.95 &      0.79 \\ 
\addlinespace[1ex] Nyabihu  &      0.96 &      0.93 &      0.91 \\ 
\addlinespace[1ex] Nyamasheke  &      0.84 &      0.81 &      0.89 \\ 
\addlinespace[1ex] Rubavu  &      0.59 &      0.93 &      0.86 \\ 
\addlinespace[1ex] Rusizi  &      0.57 &      0.91 &      0.93 \\ 
\addlinespace[1ex] Rutsiro  &      0.83 &      0.93 &      0.80 \\ 
\addlinespace[1ex] \addlinespace[1ex] 
\bottomrule
\end{tabular}

	    }
	    \end{center}

	\vskip-4ex 

	\floatfoot{
	     \textsc{Notes}---Table shows the share of teachers who remain in the same school in the subsequent year, across districts and years. Estimates based on sample of schools for which exams data are available. Estimates are average retention rates among teachers employed in each of the years 2016--2018.  Retention outcomes for Gasabo and Nyarugenge are not available for 2017 or 2018.
	}
	\end{footnotesize}
\end{table}

\end{document}